\documentclass[pre,twocolumn,showpacs]{revtex4}
\usepackage{graphicx}

\newcommand{\equ}{Eq.}
\newcommand{\eqs}{Eqs.}
\newcommand{\fig}{Fig.}
\newcommand{\figs}{Figs.}
\newcommand{\sect}{Sec.}
\newcommand{\tab}{Tab.}
\newcommand{\rem}[1]{}

\newcommand{\imag}[1]{\text{Im}(#1)}

\newcommand{\real}[1]{\text{Re}(#1)}
\newcommand{\realb}[1]{\text{Re}[#1]}

\begin{document}

\title{Hexagonal dielectric resonators and microcrystal lasers}   
\author{Jan Wiersig}
\affiliation{Max-Planck-Institut f\"ur Physik komplexer Systeme, D-01187
Dresden, Germany}
\email{jwiersig@mpipks-dresden.mpg.de}
\date{\today}

\begin{abstract} 
We study long-lived resonances (lowest-loss modes) in hexagonally shaped
dielectric resonators in order to gain insight into the physics of a class
of microcrystal lasers. 
Numerical results on resonance positions and lifetimes, near-field intensity
patterns, far-field emission patterns, and effects of rounding of
corners are presented. 
Most features are explained by a semiclassical approximation based on
pseudointegrable ray dynamics and boundary waves. The semiclassical model is 
also relevant for other microlasers of polygonal geometry.
\end{abstract}
\maketitle

\section{Introduction}
A novel class of microlasers based on nanoporous molecular
sieve host-guest systems has been fabricated recently by Vietze 
{\it et al.}~\cite{VKLISLA98}. Organic dye guest molecules were put
into the channel pores of a zeolitic microcrystal $\text{AlPO}_4-5$ 
host. The aluminophosphate-crystals grow with natural hexagonal boundaries
with a high degree of perfection. 
In terms of pump energy needed to reach lasing threshold these
microlasers can compete with semiconductor based vertical-cavity
surface-emitting lasers. This makes them a promising candidate for future
applications as e.g. optical communication devices. 

Microlasers and microresonators are not only relevant for experiments and
applications but  
they are also of great interest from a theoretical point of view 
because they are {\it open} (or leaky) and they can be {\it mesoscopic}. 
In our case the latter depends on the lasing dye and on the crystal size. Both
introduce a characteristic length scale: the wavelength $\lambda$ ranging
from $600\text{nm}$ to $800\text{nm}$, and the side length $R$ of the
hexagonal cross section of the crystal ranging from $2.6\mu\text{m}$ to 
$4.6\mu\text{m}$ in Refs.~\cite{VKLISLA98,BILNSSVWW00}. In current experiments,
larger crystals with $R$ up to $20\mu\text{m}$ are under investigation.  
Only the ratio of these two length scales is relevant. It can be
expressed by the dimensionless size parameter $20\leq kR\leq 190$, where
$k=2\pi/\lambda$ is the free-space wave number. For small $kR$ 
the wavelength and the cavity size are of the same order. The system is
microscopic. For large but finite $kR$ the system is mesoscopic.  

The theoretical analysis of the lasing modes reduces to the two-dimensional 
problem of resonant modes in a passive dielectric microresonator of regular
hexagonal geometry (ignoring surface roughness). 
In Ref.~\cite{BILNSSVWW00} preliminary numerical computations of the full wave
equations have been 
restricted to the near-field intensity pattern. We present a systematic
numerical analysis of the hexagonally shaped dielectric resonator with $20\leq
kR\leq 60$, including resonance positions and lifetimes, near-field intensity
patterns, far-field emission patterns, and effects of rounding of corners. 
To avoid convergence problems at corners, which had been a problem in
Ref.~\cite{BILNSSVWW00}, we will employ the
boundary element method (BEM)~\cite{Wiersig02b}. 

To study the deep mesoscopic regime $kR > 60$ we will introduce a
semiclassical ray model.  
The semiclassical (short-wavelength) approximation is applied in the field of 
quantum chaos to relate quantum (wave) dynamics to their underlying classical
(ray) dynamics. Most research efforts have been focused on closed resonators,
so-called billiards, where the dielectric interface is replaced by a hard wall
on which the wave function vanishes.
The classical dynamics is free motion inside the billiard with specular
reflections at the boundary.

The regular hexagon belongs to the class of rational polygonal billiards.
All angles $\phi_j$ between sides are rationally related to $\pi$,
i.e. $\phi_j = m_j\pi/n_j$, where $m_j, n_j > 0$ are relatively prime 
integers. If all $m_j$ are equal to unity, for example in the case of the 
rectangle, then the dynamics is integrable. In the presence of {\it critical
corners} with $m_j > 1$ the dynamics is not integrable but instead {\it
pseudointegrable}~\cite{RichensBerry81}. As for integrable systems the phase 
space is foliated by two-dimensional invariant 
surfaces~\cite{Hobson75,ZemlyakovKatok75}. 
However, there are some peculiar features that distinguish these billiards
strongly from integrable ones (and also from fully or partially chaotic
billiards):
(i) an invariant surface does not have the topology of a torus but instead
that of a surface of higher genus~\cite{RichensBerry81}, loosely 
speaking, a torus with additional handles;
(ii) the dynamics is not quasiperiodic. This is reflected, for example, by 
multifractal Fourier spectra of classical observables~\cite{AGR00,Wiersig00};
(iii) the quantum-classical correspondence is exotic~\cite{Wiersig01}; 
(iv) the quantum spectrum obeys critical statistics~\cite{BGS99}.
To compute the quantum spectrum of a pseudointegrable billiard with a 
semiclassical treatment
is extremely difficult, if possible at all. In the present paper, we will
demonstrate that for a {\it sufficiently open} hexagonal dielectric resonator 
the spectrum and the eigenmodes can be computed semiclassically. 
As a byproduct, we gain an intuitive understanding of the numerical results.

The paper is organized as follows. The system is defined in
\sect~\ref{sec:system}. Section~\ref{sec:numerics} presents the numerical
analysis. The semiclassical ray model is introduced in
\sect~\ref{sec:raymodel}. Finally, \sect~\ref{sec:conclusion} contains our
conclusions.  

\section{The System}
\label{sec:system}
The numerical computation of lasing modes in a cavity with active medium and
complex geometry is an extremely difficult task; see 
Refs.~\cite{HDI99,HDI02}. Focusing on the effects of the geometry we here
restrict our calculations to resonant modes in a passive cavity. Which of
these resonant modes contribute to lasing is not important in our case since
all long-lived resonant modes have very similar properties as we will see in
the following. 

In the experiments on the microcrystals it has been shown that the 
electromagnetic field is TM polarized~\cite{VKLISLA98,BILNSSVWW00}. Maxwell's
equations reduce to a two-dimensional wave equation~\cite{Jackson83}    
\begin{equation}\label{eq:wave}
-\nabla^2\psi = n^2({\bf r})k^2\psi \ ,
\end{equation}
with wave number $k$ and piece-wise constant index of refraction $n({\bf
r})$. The index  
of refraction is $n=1.466$ inside the cavity and $1$ outside. The origin of
the coordinate system ${\bf r} = (x,y) = (r\cos{\theta},r\sin{\theta})$ is
located in the center of the hexagonal cavity. 
The complex-valued wave function $\psi$ represents the 
$z$-component of the real-valued electric field vector $E_z({\bf r},t) = \realb{\psi({\bf
r})\exp{(-i\omega t)}}$ with $i^2=-1$, 
angular frequency $\omega=ck$ and speed of light in vacuum $c$.
$\psi$ and $\nabla\psi$ are continuous at the boundary of the resonator.
To model the situation in a laser, we impose the outgoing-wave condition
\begin{equation}\label{eq:outgoingbc}
\psi \sim h(\theta,k)\frac{\exp{(ikr)}}{\sqrt{r}} 
\end{equation}
for large $r$.
With a real-valued $n$ (passive resonator without absorption) this leads to
modes that are exponentially decaying in time. 
The lifetime $\tau$ of these so-called resonant modes or short {\it
resonances} is given by $\tau=-1/2c\,\imag{k}$ with $\imag{k} < 0$. The 
lifetime $\tau$ is related to the quality factor $Q = \real{\omega}\tau$. 
The resonant modes are connected to the peak 
structure in scattering spectra; see e.g. Ref.~\cite{Landau96}. 
We are here only interested in long-lived resonances that provide a
sufficiently long lifetime for the light to accumulate the gain required to
overcome the lasing threshold. Note that extremely long-lived
resonances are not relevant for lasers because they do not supply enough
output power.  

\subsection{Symmetry considerations}
\begin{figure}[ht]
\includegraphics[width=8.0cm,angle=0]{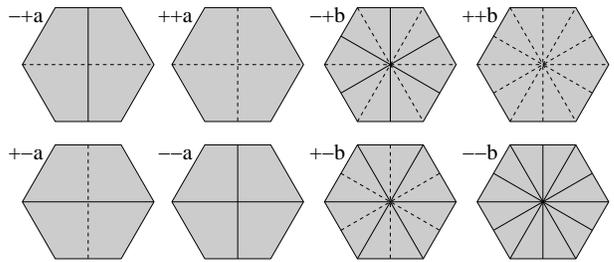}
\caption[]{\footnotesize The eight symmetry classes of the hexagon. Odd (Even)
symmetry is marked by solid (dashed) lines.} 
\label{fig:symmetries}
\end{figure}

Figure~\ref{fig:symmetries} shows the eight symmetry classes of the hexagon;
see e.g. Ref.~\cite{CK99}. In the notation $-+a, \ldots$, the first sign is 
$+$ if the wave function is even with respect to $x\to-x$, and $-$ otherwise. 
Correspondingly, the second sign refers to $y\to-y$. The letter $a$ indicates
two symmetry lines, whereas the letter $b$ indicates six symmetry lines. 

The $+-a$ and the $-+a$-modes have exactly the same complex $k$. 
The reason is that a $+-a$-mode can be converted to a $-+a$-mode and vice
versa in the following way. Take two copies of a single $+-a$-mode. Rotate the
first copy by $60^\circ$ clockwise around the origin of the coordinate system
and the second copy by the same angle counterclockwise. Subtracting the two
gives a $-+a$-mode with the same $k$. The way from $-+a$ to $+-a$, from $++a$
to $--a$, and from $--a$ to $++a$ is analogue. Hence, $a$-modes always appear 
in degenerate pairs having the same $k$.

Each linear superposition of such a pair of degenerate $a$-modes, we denote
them by $\psi_1$ and $\psi_2$, is also a solution of the wave
equation~(\ref{eq:wave}). Because of the 6-fold symmetry of the system, we can
find always two superpositions $\psi_\pm = \psi_1+p_\pm\psi_2$
with the property that the corresponding intensities $|\psi_+|^2$ and
$|\psi_-|^2$ are invariant under $60^\circ$-rotations. That means $\psi_+$ and
$\psi_-$ change only by a phase factor $\exp{i\Omega_\pm}$ when rotated in
real space by $60^\circ$. A full rotation about $360^\circ$ does not change
the wave function. Hence, $\Omega_\pm = \pi q/3$ with $q=-2,-1,0,1,2,3$. Yet,
$q=0$ and $q=3$ 
are not allowed because the intensities of $a$-modes are not invariant under
$60^\circ$-rotations. A rotation about $180^\circ$ is identical to
$(x,y)\to(-x,-y)$. From this we find $\Omega_\pm = \pm\pi/3$ for $\psi_\pm$ formed
by $a$-modes of type $+-$, $-+$ and $\Omega_\pm = \pm 2 \pi/3$ for $\psi_\pm$
formed by $a$-modes of type $++$, $--$. These requirements lead to 
\begin{equation}\label{eq:p}
p_\pm = -\frac{\exp{(i\Omega_\pm)} \psi_1({\bf r})-\psi_1(\tilde{{\bf r}})} 
              {\exp{(i\Omega_\pm)} \psi_2({\bf r})-\psi_2(\tilde{{\bf r}})} 
\end{equation}
where ${\bf r}\neq (0,0)$ is an arbitrary point and $\tilde{{\bf r}}$ is ${\bf
r}$ rotated by $60^\circ$ counterclockwise. Formula~(\ref{eq:p}) is valid for
any relative phase of $\psi_1$ and $\psi_2$.

The $b$-modes cannot be converted into each other since each $b$-mode is 
invariant under $60^\circ$-rotations up to a phase. Hence, $b$-modes do not
form degenerate pairs. 

\subsection{Rounding}
\begin{figure}[ht]
\centerline{
a)\includegraphics[width=3.5cm,height=3.1cm,angle=0]{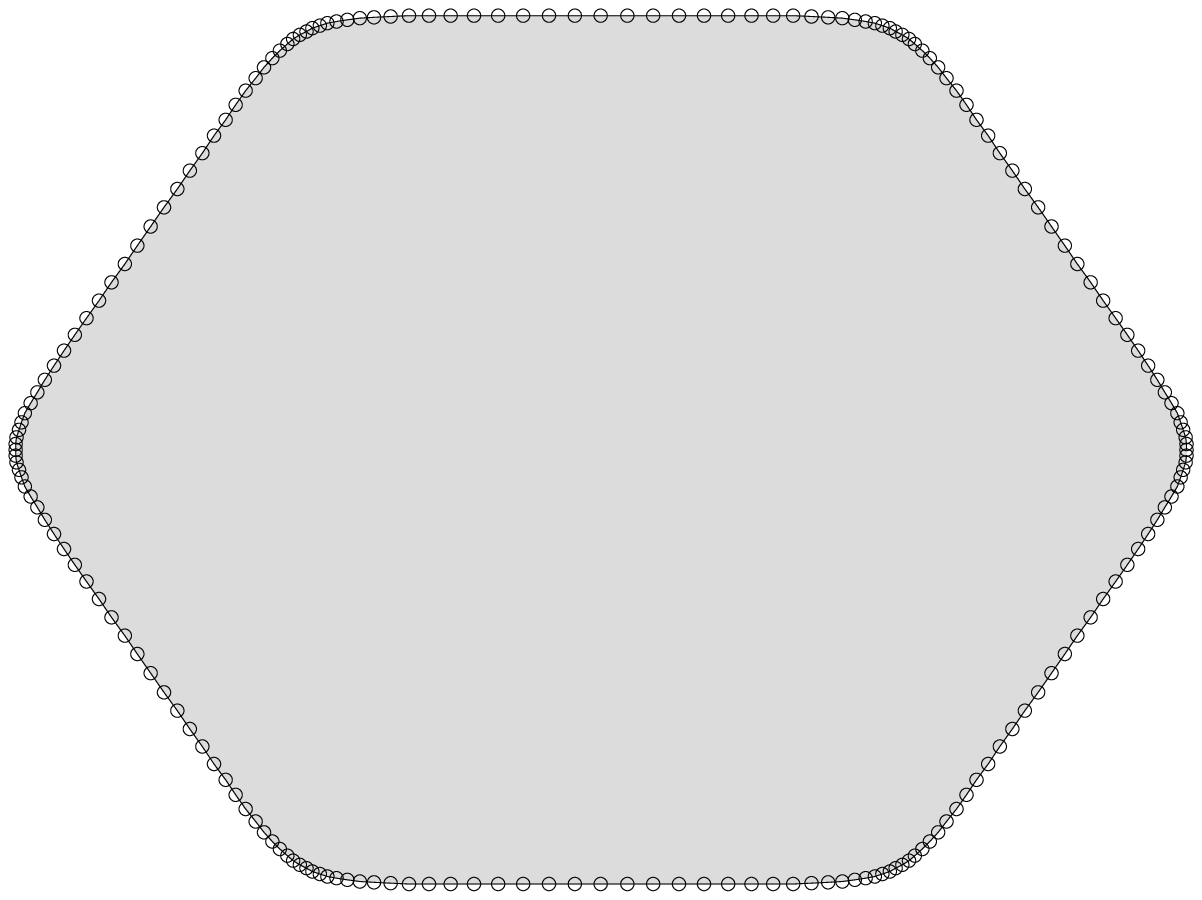}
b)\includegraphics[width=3.5cm,height=3.1cm,angle=0]{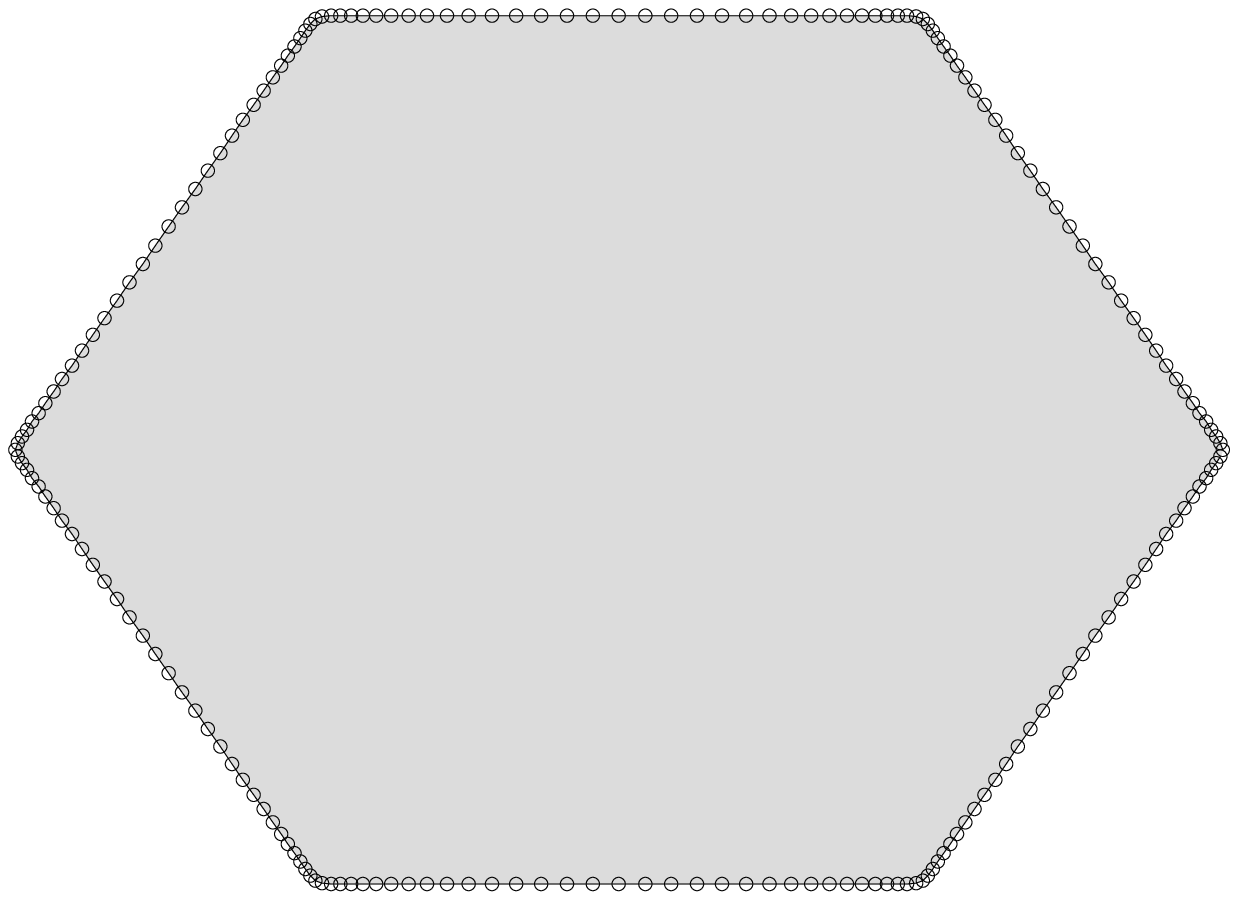}
}
\caption[]{\footnotesize Shape and discretization (circles) of 
the resonator with (a) s = 20, (b) s = 200 and $\eta = 0.1$. For
illustrational reasons only $N = 200$ discretization points are used.} 
\label{fig:para}
\end{figure}

For our numerics it will be necessary to round the corners of the resonator 
slightly as depicted in \fig~\ref{fig:para}. In terms of polar coordinates 
the parametrization of the boundary reads 
\begin{equation}\label{eq:para}
r^s = \frac{2R^s}{(\cos{\theta}-\frac{1}{\sqrt{3}}\sin{\theta})^s+(\frac{2}{\sqrt{3}}\sin{\theta})^s+(\cos{\theta}+\frac{1}{\sqrt{3}}\sin{\theta})^s}
\end{equation}
where $s$ is the rounding parameter, a positive even integer. $s=2$ gives the
circle, whereas $s\to\infty$ gives the hexagon with flat sides and sharp
corners. The parametrization in \equ~(\ref{eq:para}) preserves the full
symmetry of the problem. Note that the parametrization used in
Ref.~\cite{BILNSSVWW00} is different. 

It is easy to show that the maximum curvature for the rounded hexagon in
\equ~(\ref{eq:para}) is given by $(s-1)/3R$ for $s 
\geq 6$. The radius of curvature $\rho$ per wavelength $\lambda =
2\pi/\real{k}$ is therefore given by 
\begin{equation}
\frac{\rho}{\lambda} = \frac{3}{2\pi}\frac{\real{kR}}{s} 
\end{equation}
for large $s$. 

\section{Numerical analysis}
\label{sec:numerics}
A frequently used numerical method to solve wave equation~(\ref{eq:wave})
with outgoing-wave condition~(\ref{eq:outgoingbc}) is the wave- or
mode-matching method~\cite{ND95}. The wave function is expanded in
integer Bessel functions inside the cavity and in Hankel functions of first
kind outside, so that the outgoing-wave condition~(\ref{eq:outgoingbc}) is
fulfilled automatically.  
The Rayleigh hypothesis asserts that such an expansion is always
possible. However, it can fail for geometries which are not sufficiently weak
deformations of a circular cavity~\cite{BergFokkema79}.  
So, using the mode-matching method for the hexagonal resonator in
Ref.~\cite{BILNSSVWW00} is questionable. And indeed, according 
to Ref.~\cite{Noeckelpc02}, serious convergence problems appear in this
situation. To avoid this problem we apply the BEM~\cite{Wiersig02b}.  
Exploiting Green's identity, the two-dimensional differential 
equation~(\ref{eq:wave}) is replaced by a one-dimensional integral equation 
defined only along the boundary of the cavity. To be slightly more precise, 
the boundary appears twice in the formalism: one bounds the interior of the
cavity and one bounds the exterior. 
The boundaries are then discretized by dividing them into $2N$ small boundary 
elements. For the boundary~(\ref{eq:para}) we choose the following 
discretization 
\begin{equation}
\theta = \xi+\eta\sin{[6(\xi+\pi/2)]} \ .
\end{equation}
The discretization points on the circle $\xi\in[0,2\pi)$ are chosen to
be spaced equidistantly. The parameter $\eta$ determines the distribution of 
points on the cavity (the shape of which is completely determined by the
rounding parameter $s$). $\eta = 0$ gives an uniform density of
points while $\eta > 0$ gives an enhancement near the corners as illustrated in
\fig~\ref{fig:para}. We always use $\eta = 0.1$. 

\subsection{Resonance positions in the complex plane}
\begin{figure}[ht]
\includegraphics[width=6.0cm,angle=0]{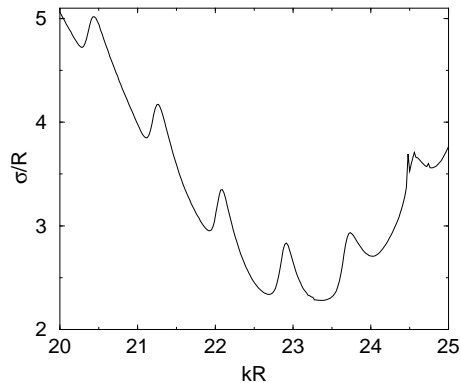}
\caption[]{\footnotesize Calculated total scattering cross section $\sigma/R$ vs. $kR$ for
a hexagonal resonator. The plane wave is incidence at $15^\circ$ to
the horizontal side faces. $s=100$, and $2N = 2000$.} 
\label{fig:sigma}
\end{figure}

Figure~\ref{fig:sigma} shows the total cross section $\sigma$ as function of
the dimensionless wave number $kR$ [$\imag{kR} = 0$] for plane-wave 
scattering with incidence angle $\theta = 15^\circ$. For numerical details of
how to compute this quantity see Ref.~\cite{Wiersig02b}. We observe a {\it
single-mode spectrum}, a series of roughly equidistant peaks 
(the fine structure around $kR \approx 24.5$ is an artifact of the
BEM which can be removed by increasing the number of
discretisation points~\cite{Wiersig02b}).  
We will count the peaks by the {\it mode index} (or quantum number) $m$
according to increasing $kR$ starting with $m=0$ at $kR=0$. The first peak at
$kR \approx 20.5$ in \fig~\ref{fig:sigma} is then labeled by $m=23$. 
The position of a given peak $kR$ and its width $\Gamma$ in
\fig~\ref{fig:sigma} are related to the complex value of $kR_{\text{mode}}$ of
the corresponding resonant mode by means of $kR_{\text{mode}} \approx
kR-i\Gamma/2$. The subscript ``mode'' will be dropped from now on.    

\begin{figure}[ht]
\includegraphics[width=6.5cm,angle=0]{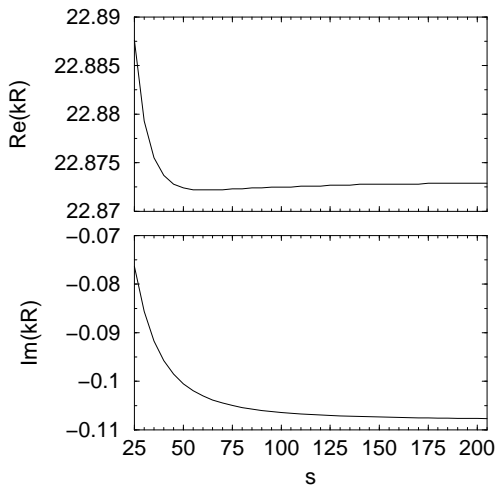}
\caption[]{\footnotesize Real and imaginary part of $kR$ as function of the
rounding parameter $s$. $2N = 2000$, $m=26$, $kR = 22.8725 -i0.1064$ for $s=100$.}
\label{fig:dynamicsrounding}
\end{figure}

The BEM not only solves the scattering problem but it also can compute 
the resonant modes. In \fig~\ref{fig:dynamicsrounding} we plot the real and
imaginary part of $kR$ of 
a resonance as function of the rounding parameter $s$. $\real{kR}$ saturates 
around $s=55$, whereas $\imag{kR}$ saturates at
$s\geq100$. We take $s=100$ translating to $\rho/\lambda \approx 0.11$. 
It is surprising that one has to decrease $\rho$ to values one order of
magnitude smaller than the wavelength. In the following we will fix
$\rho/\lambda \approx 0.11$, that means when we change $\real{kR}$, we have to
change $s$ accordingly.   

\begin{figure}[ht]
\includegraphics[width=7.5cm,angle=0]{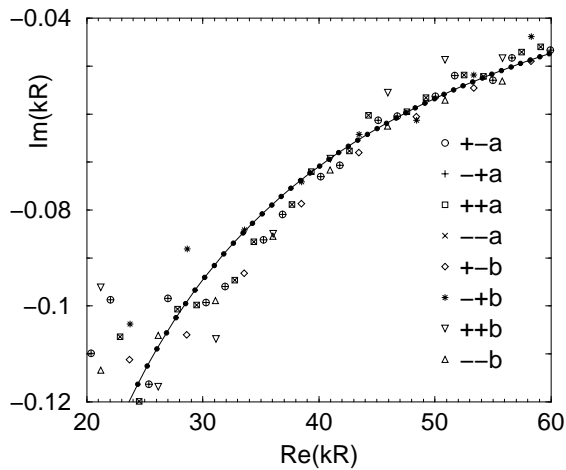}
\caption[]{\footnotesize Long-lived resonances in the complex plane,
cf. \fig~\ref{fig:symmetries}. Filled circles on the solid line are the
semiclassical solutions in \eqs~(\ref{eq:qcondition}) and (\ref{eq:kikr}) from
\sect~\ref{sec:raymodel}. 
} 
\label{fig:complexplane} 
\end{figure}

Figure~\ref{fig:complexplane} shows the long-lived resonances in the complex
plane inside the strip $20 \leq \real{kR} \leq 60$. Our numerics cannot cover 
the full regime of the microlaser experiments~\cite{VKLISLA98,BILNSSVWW00} $20
\leq \real{kR} \leq 190$. Small values of $|\imag{kR}|$ correspond to
long-lived modes and, correspondingly, large values correspond to short-lived
modes.  
Four features in \fig~\ref{fig:complexplane} are striking: 
(i) the $a$-modes are two-fold degenerated as predicted from the symmetry
considerations. 
(ii) The $b$-modes come in quasi-degenerated pairs with slightly different 
$kR$ (differences in $\real{kR}$ and $\imag{kR}$ are of the same
order which cannot be seen in \fig~\ref{fig:complexplane}). It is not possible
to resolve these small splittings in the scattering cross section in 
\fig~\ref{fig:sigma}.  
(iii) The values of $\real{kR}$ are approximately equidistantly spaced. 
(iv) Highly-excited resonances with $\real{kR} \geq
34$ lie approximately on a smooth curve. 

\begin{figure}[ht]
\includegraphics[width=6.5cm,angle=0]{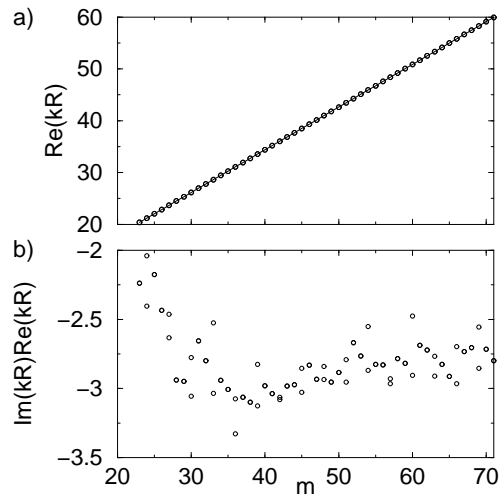}
\caption[]{\footnotesize (a) Wave number $\real{kR}$ vs. mode index $m$. The
data can be approximated by a straight line $\real{kR} = \nu(m+m_0)$ with
$\nu\approx 0.8238$ and $m_0\approx 1.7516$.
(b) $\imag{kR}\,\real{kR}$ vs. $m$. For $m\geq 34$, the data is well
approximated by the mean value $\approx -2.87$.} 
\label{fig:modes} 
\end{figure}

The equidistant spacing of $\real{kR}$ can be seen more clearly in 
\fig~\ref{fig:modes}(a). Pairs of (quasi-) degenerated modes are labeled by
the mode index. We see that all pairs lie extremely close to a line
$\real{kR} = \nu(m+m_0)$ with mode spacing $\nu$ and shift $m_0$. Linear 
regression gives $\nu\approx 0.8238$ and $m_0\approx 1.7516$. The same mode
spacing can be observed in \fig~\ref{fig:sigma}. The corresponding free
spectral range $\Delta\lambda = \nu\lambda^2/2\pi R$ is in agreement with the
experiments~~\cite{VKLISLA98,BILNSSVWW00}.

From \fig~\ref{fig:modes}(b) we can infer that the product of real and
imaginary part of $kR$ for highly-excited modes does not depend on the mode
index $m$. We find  
\begin{equation}\label{eq:hyperbola}
\imag{kR}\real{kR} \approx -2.87 
\end{equation}
for $n=1.466$. Later we will see that this remarkable relation is a good
approximation not only in the numerical accessible regime but also in the full
experimental regime. Relation~(\ref{eq:hyperbola}) implies that the lifetime is
proportional to $\real{k}R^2$ and the quality $Q$ is proportional to
$\real{kR}^2$ ($Q$ ranges from roughly $70$ to $6400$). This finding is
relevant for the future experiments on microcrystal lasers: it implies
that the laser threshold decreases as the size of the resonator is increased.

Due to the qualitative change in behaviour around $\real{kR} \approx 34$ in
\figs~\ref{fig:complexplane} and \ref{fig:modes}(b) we distinguish the regions
$\real{kR} < 34$ and $\real{kR} > 34$. We refer to the
former one as the microscopic regime and to the latter as the mesoscopic
regime.

\subsection{Mode structure}
\begin{figure}[ht]
\includegraphics[width=8.5cm,angle=0]{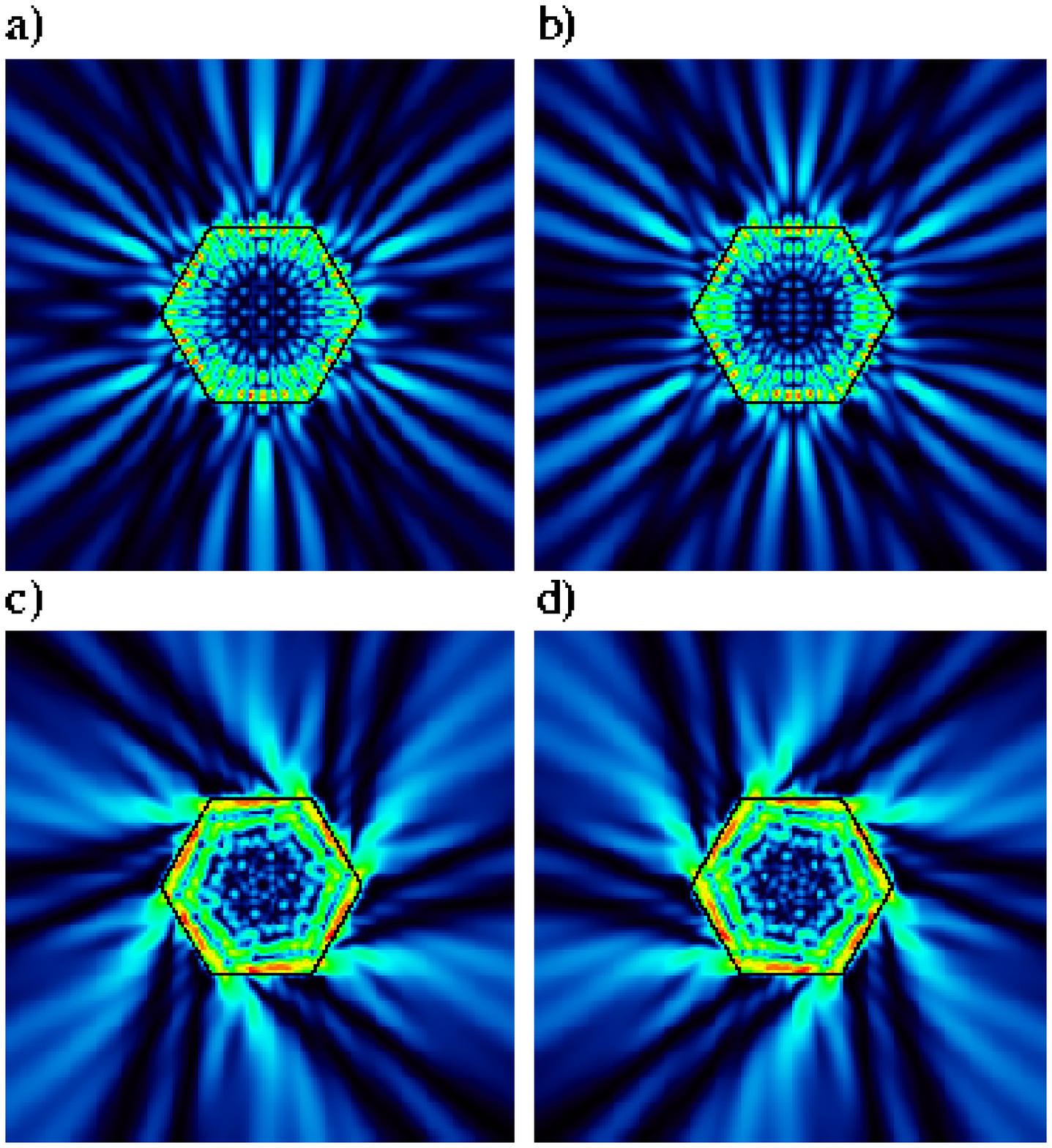}
\centerline{
\includegraphics[width=8.5cm,angle=0]{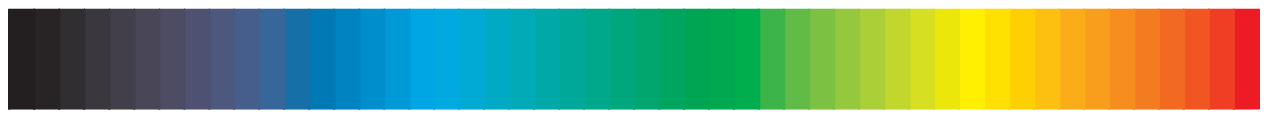}
}
\caption[]{\footnotesize Calculated near-field intensity pattern $|\psi({\bf
r})|^2$ of two-fold degenerated resonances. (a) $26++a$ and (b) $26--a$ with
well-defined parity. (c) $26+$ and (d) $26-$ with well-defined chirality. 
Intensity is higher for redder colours, and vanishes in the dark regions. $kR
= 22.8725 -i0.1064$, $s=100$, $2N = 2000$.} 
\label{fig:lowresonance}
\end{figure}

Figures~\ref{fig:lowresonance}(a) and (b) show the near-field intensity
pattern of the resonances $26++a$ and $26--a$ (mode index $m=26$), respectively. While 
these two resonances are standing waves (ignoring the uniform temporal decay), 
the corresponding superpositions $\psi_+$ and $\psi_-$ in
\figs~\ref{fig:lowresonance}(c) and (d) are unidirectional traveling waves. 
We call them chiral modes. Such a mode is specified by its mode index $m$ and
a label $+$ (traveling counterclockwise) or $-$ (traveling clockwise). In the
following we will deal only with these chiral modes 
which are more straightforward to compare to the ray dynamics.
We remark that the Husimi function representation frequently applied to 
optical microcavities (see e.g. Ref.~\cite{HSS02}) does not provide more 
insight in our case.  

\begin{figure}[ht]
\includegraphics[width=8.0cm,angle=0]{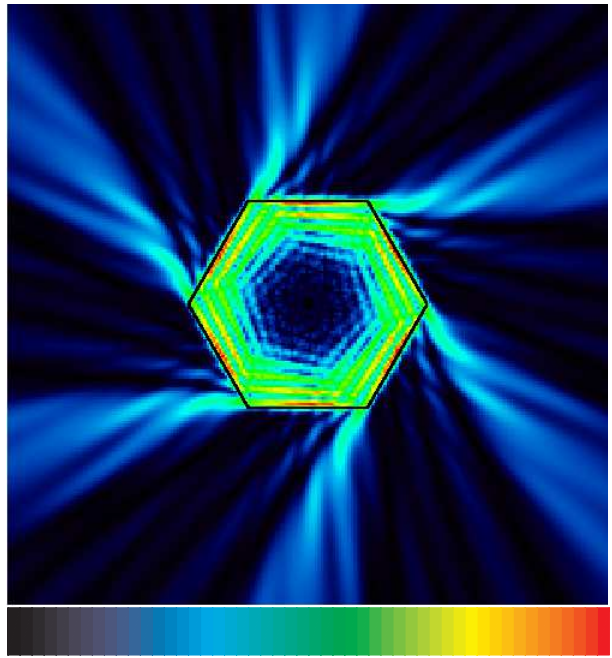}
\centerline{
\includegraphics[width=8.0cm,angle=0]{colormap.eps}
}
\caption[]{\footnotesize Chiral resonance $50-$. $kR = 42.6318-i0.06766$, 
$s=200$, $2N = 4000$.}  
\label{fig:highresonance} 
\end{figure}

\begin{figure}[ht]
\includegraphics[width=8.0cm,angle=0]{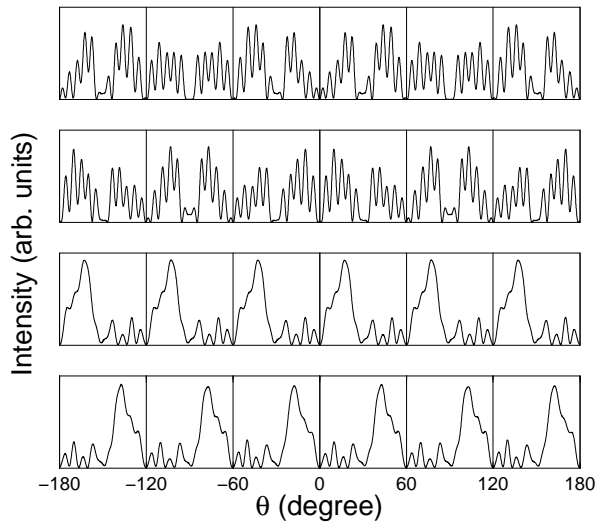}
\caption[]{\footnotesize Far-field emission pattern, $|\psi(r,\theta)|^2$ for 
large $r$, of (from above) $50--a$, $50++a$, $50-$, and $50+$
modes; cf. \fig~\ref{fig:highresonance}. Vertical lines mark the directions
parallel to the edges.} 
\label{fig:farfield} 
\end{figure}

Figure~\ref{fig:highresonance} shows a higher-excited superposition $\psi_-$. 
The following properties can be observed:
(i) the intensity is concentrated along the boundary of the cavity, resembling
{\it whispering-gallery modes} in circular or weakly deformed circular
cavities. 
(ii) The wave pattern looks regular. An 
approximate nodal-line structure with a peculiar twist is visible.
(iii) The {\it emission is predominantly at the corners}. This is in agreement
with the laser emission measured in the experiments in
Refs.~\cite{VKLISLA98,BILNSSVWW00}. (iv) Outside the 
cavity the light propagates along certain directions {\it not parallel to the
facets}. The latter fact can be seen more clearly in the far-field distribution
shown in \fig~\ref{fig:farfield}. There are six emission peaks with angular
width of $\approx 14^\circ$ and angular distance to the nearest facet of 
$\approx 17^\circ$.

\begin{figure}[ht]
\includegraphics[width=8.0cm,angle=0]{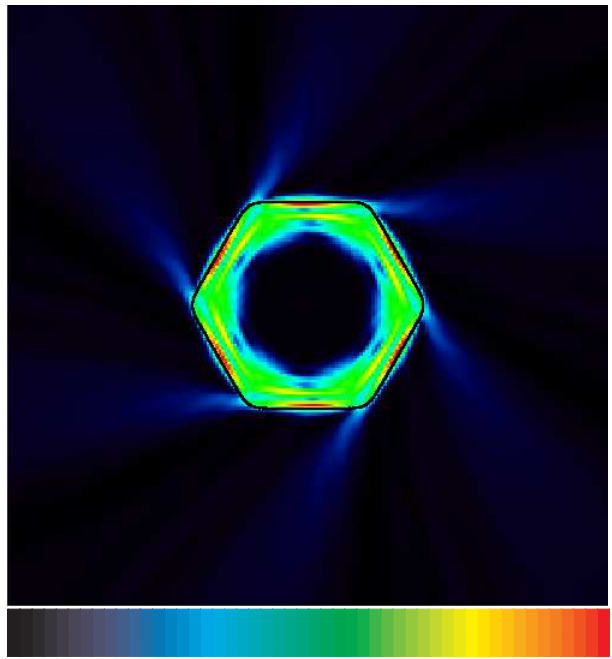}
\centerline{
\includegraphics[width=8.0cm,angle=0]{colormap.eps}
}
\caption[]{\footnotesize Resonance $50-$ in a rounded hexagon with $s=20$,
cf. \fig~\ref{fig:highresonance}. $kR = 42.7099-i0.01836$, $2N = 4000$.}  
\label{fig:roundednear} 
\end{figure}

\begin{figure}[ht]
\includegraphics[width=8.0cm,angle=0]{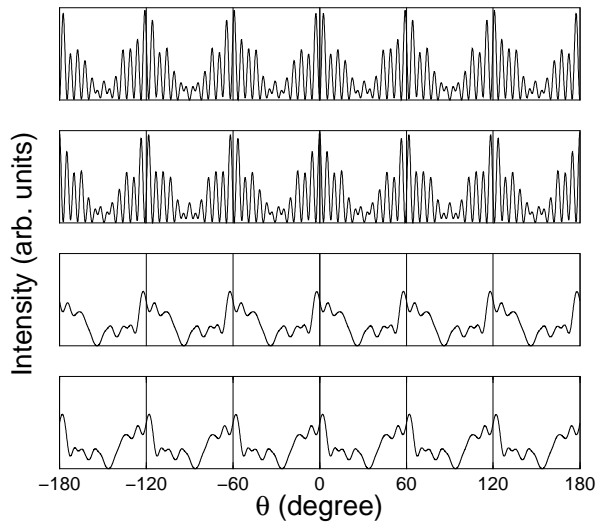}
\caption[]{\footnotesize Far-field emission pattern of the resonance (from
above) $50--a$, $50++a$, $50-$, and $50+$ in the
rounded hexagon; cf. \fig~\ref{fig:roundednear}.}  
\label{fig:roundedfar} 
\end{figure}

The {\it sensitivity to rounding} found for the resonance positions in the 
complex plane, see \fig~\ref{fig:dynamicsrounding}, also
shows up in the mode structure. Figure~\ref{fig:roundednear} shows the
near-field intensity pattern of a
rounded hexagon with $s=20$, i.e. $\rho \approx \lambda$. The emission is
again at the corners, but it is reduced. Moreover, the directionality has
decreased and the high-intensity directions are now parallel to the edges. 
The latter fact can be seen better in the far-field emission pattern in
\fig~\ref{fig:roundedfar}.  

\section{Semiclassical analysis}
\label{sec:raymodel}
Having computed numerically the long-lived resonances in the regime $20\leq
\real{kR}\leq 60$, we now introduce a semiclassical ray model in order to see
what happens in the deep mesoscopic regime $\real{kR} > 60$. 
So far, semiclassical approximations of dielectric resonators have been
discussed only for the case of smooth boundaries~\cite{ND95,ND97,NHJS99}. 
Our heuristic approach is divided into three steps: geometric optics,
semiclassical quantization and emission mechanisms.  

\subsection{Geometric optics}
\begin{figure}[ht]
\includegraphics[width=5.0cm,height=5.3cm,angle=0]{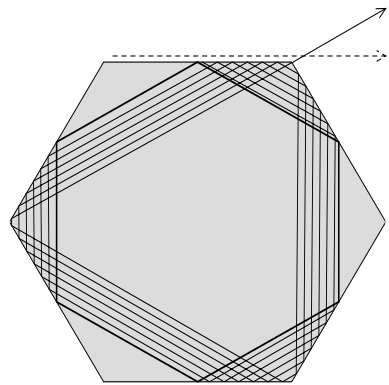}
\caption{\footnotesize Semiclassical ray model. Thick line marks a member of 
the family of long-lived rays, other members are obtained by shifting
the ray along the boundary (not shown). The thin line marks a ray with
slightly different angle of incidence. Arrows indicate emission due to
pseudointegrable dynamics (thin) and boundary waves (dashed).} 
\label{fig:raymodel}
\end{figure}

Geometric optics follows from wave equation~(\ref{eq:wave})
in the limit $\real{kR}\to\infty$ neglecting all interference effects.
In the following we focus on low-index materials with 
$n_{\text{min}} < n < n_{\text{max}}$ where $n_{\text{min}} = 1/\sin{60^\circ}
\approx 1.16$ and $n_{\text{max}} = 1/\sin{30^\circ} = 2$. The lower bound
guarantees that a six-bounce periodic ray with angle of incidence
$\theta_i = 60^\circ > \theta_c$ (with $\sin\theta_c=1/n$) is trapped within
the hexagonal resonator by total internal reflection at the facets; see
\figs~\ref{fig:raymodel} (some aspects of the much
simpler case $n < n_{\text{min}}$ have been studied in 
Ref.~\cite{Bhowmik00}). The periodic ray is marginally stable with respect to
shifting it along the boundary. In this way we obtain a whole family of
periodic rays with identical length and angle of incidence. 

The upper bound $n_{\text{max}} = 2$ ensures that triangular-shaped periodic
rays with $30^\circ$ angle of incidence are not totally reflected. The
periodic-ray family with $\theta_i = 60^\circ$ is then the only long-lived
family of short period. Yet, since periodic rays are dense in phase space,
we cannot exclude the possibility that there are also long-lived periodic
rays of high period. 

It should be emphasized at this point that the (infinite) long lifetime of the
rays is relevant for our purpose, whereas the periodicity of the rays is not relevant. 
The latter comes in here simply because periodic rays (and their neighborhood)
can be long-lived. We will not apply methods from semiclassical periodic-orbit 
theory~\cite{Gutz90}. 

The periodic rays are unstable with respect of changing the angle of
incidence. Figure~\ref{fig:raymodel} shows a ray with slightly different
initial angle of incidence. The ray is slowly diverging from the central one.
After some time, it reaches the corner on its other side at (almost) normal
incidence. Consequently, it then escapes refractively (with probability close
to 1). 
\rem{
\begin{figure}[ht]
\includegraphics[width=7.0cm,angle=0]{interface.eps}
\caption{\footnotesize Reflection and refraction at a dielectric interface.}
\label{fig:interface}
\end{figure}
}

\begin{figure}[ht]
\includegraphics[width=8.0cm,angle=0]{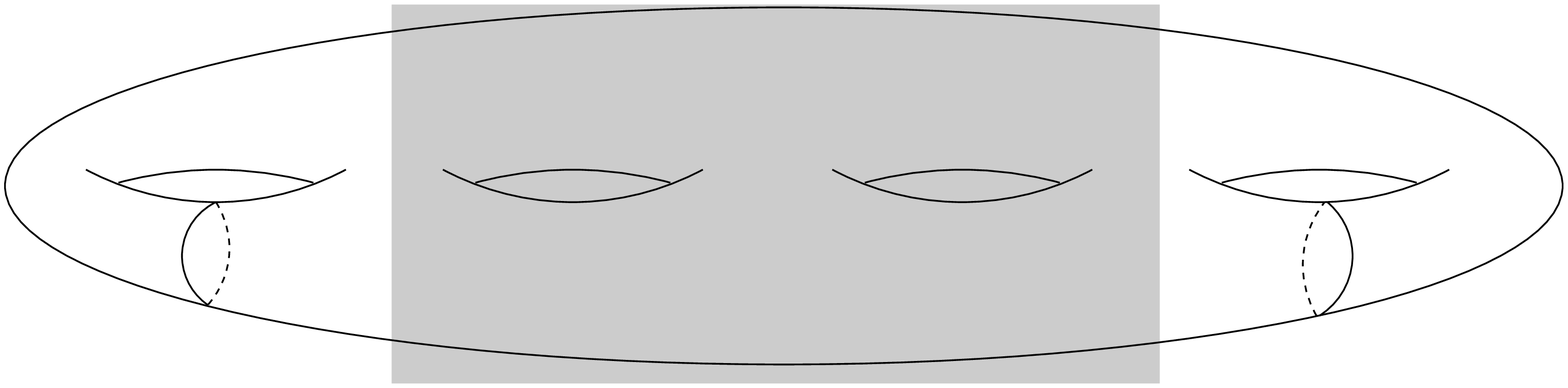}
\caption{\footnotesize Invariant surface of the hexagonal billiard. The small
circle on the left (right) hand side represents the central periodic ray in
\fig~\ref{fig:raymodel} cycling (counter-) clockwise. 
In the grey region the condition for total internal reflection is not 
fulfilled.} 
\label{fig:torus4}
\end{figure}

As long as the ray does not escape from the cavity it behaves as if it were in
a closed resonator, i.e. in a billiard. It is illuminating to examine the ray
dynamics of the open system in terms of the invariant surface of the 
hexagonal billiard. The genus of the invariant surfaces is given by the
general formula~\cite{RichensBerry81} 
\begin{equation}
g = 1+\frac{{\cal N}}{2}\sum_j\frac{m_j-1}{n_j}
\end{equation}
where the sum is over all corners with angles $\phi_j = m_j\pi/n_j$ and 
${\cal N}$ is the least common multiple of the $n_i$. For the hexagon $m_j =
2$ and $n_j = 3$ for all $j$; hence, ${\cal N} = 3$ and finally $g = 4$. 
The surface of genus four is shown schematically in \fig~\ref{fig:torus4}. The
topology is the same for all initial conditions. First, we consider
the special case where the surface is foliated by periodic orbits. 
Each periodic ray with $\theta_i = 60^\circ$ appears here as two disjoint 
circles with different sense of rotation. These two circles correspond to the
two unidirectional traveling waves. In the following we will focus on the
circle with clockwise rotational sense.
On the same invariant surface we find also periodic rays of ``bouncing-ball'' 
type with angle of incidence $\theta_i = 0^\circ$. In the open system, the
bouncing-ball rays are not long-lived since they are not totally reflected. 
 
The dynamics on a generic invariant surface in the neighborhood of the
special surface discussed above is as follows. A nonperiodic ray starting near
the central periodic ray with slightly different initial angle of
incidence stays in the vicinity of the central ray by winding 
around the handle many times. Finally, it has to leave the vicinity of the 
periodic ray since pseudointegrable motion is ergodic on generic invariant 
surfaces~\cite{Gutkin96}. Before the ray can reach another handle (in the grey
region in \fig~\ref{fig:torus4}) and perform complicated dynamics, it escapes 
refractively from the cavity. Hence, the sufficiently open resonator
does not see the complicated topology of the full invariant surface but just
two disjoint tori. In this sense, the openness moves the system closer to
integrability.     
That is the reason why we can derive in the following subsection a practical 
semiclassical
approximation for the open hexagonal resonator whereas this is impossible for
the hexagonal billiard. 

\subsection{Semiclassical quantization}
Having identified the long-lived part of classical phase space, we now
quantize it in a semiclassical approximation. This approximation is valid for 
small $1/\real{kR}$ and it does include interference effects, in contrast to
the geometric-optics limit.  

Because of the absence of stability of the periodic rays the conventional
approach in laser optics, the paraxial approximation (see e.g. Ref.~\cite{TSSN02}),
does not work. Nevertheless, the semiclassical quantization is simple. The
idea is that an integer number 
$m=1,2,3,\ldots$ of wavelengths $\lambda_{\text{inside}} = \lambda/n =
2\pi/n\real{k}$ fits on the path length of the long-lived ray shown 
in \fig~\ref{fig:raymodel},
taking into account the phase shifts at the dielectric boundary. It will turn
out that the number $m$ defined in this way is identical to the mode index
$m$ as used in \sect~\ref{sec:numerics}. 

To compute the phase shifts we consider the text-book treatment of reflection
of a plane wave at an infinitely extended dielectric interface; see
e.g. Ref.~\cite{BornWolf59}. This 
simplified setup is justified if $\lambda \ll R$, i.e. $\real{kR} \gg 1$.
We shift the origin of the coordinate system such that $y=0$ is the
dielectric boundary bounding the lower-index region from below and the
higher-index region from above.   
In the higher-index region there is an incident wave 
\begin{equation}\label{eq:waveinc}
\psi_{i} = A_i\exp{[i\real{k}n(x\sin{\theta_i}+y\cos{\theta_i})]} \ ,
\end{equation}
with amplitude $A_i$. We set $A_i =1 $ without loss of generality. The
reflected wave is given by
\begin{equation}\label{eq:waveref}
\psi_{r} = A_r\exp{[i\real{k}n(x\sin{\theta_r}-y\cos{\theta_r})]} \ .
\end{equation}
In the lower-index region there is an emitted wave
\begin{equation}\label{eq:evanescent}
\psi_{e} = A_e\exp{[i\real{k}(x\sin{\theta_e}+y\cos{\theta_e})]} \ .
\end{equation}
The boundary conditions
\begin{equation}
\psi_i(x,0)+\psi_r(x,0) = \psi_e(x,0)
\end{equation}
and
\begin{equation}
\frac{\partial\psi_i}{\partial
y}\Biggl|_{(x,0)}+\frac{\partial\psi_r}{\partial y}\Biggl|_{(x,0)} = \frac{\partial\psi_e}{\partial y}\Biggl|_{(x,0)}
\end{equation}
lead to $\theta_i = \theta_r$, Snell's law $n\sin\theta_i =
\sin\theta_e$, $A_i+A_r=A_e$ and the Fresnel formula
\begin{equation}
A_r = \frac{1-i\alpha}{1+i\alpha}
\end{equation}
with 
\begin{equation}\label{eq:alpha}
\alpha = \frac{\sqrt{n^2\sin^2\theta_i-1}}{n\cos\theta_i} \ .
\end{equation} 

The quantization (or resonance) condition then reads
\begin{equation}
e^{in\real{k}l}A_r^6 = 1 \ ,
\end{equation}
with the length of the periodic rays
$l=3\sqrt{3}R$. 
After some algebraic manipulations we arrive at 
\begin{equation}\label{eq:qcondition}
\real{kR} = \frac{2\pi}{3\sqrt{3}n}(m+\beta) 
\end{equation}
with the total boundary phase shift $\beta$ given by
\begin{equation}
\tan{\frac{\pi}6\beta} = \alpha \ .
\end{equation}

\begin{figure}[ht]
\includegraphics[width=7.5cm,angle=0]{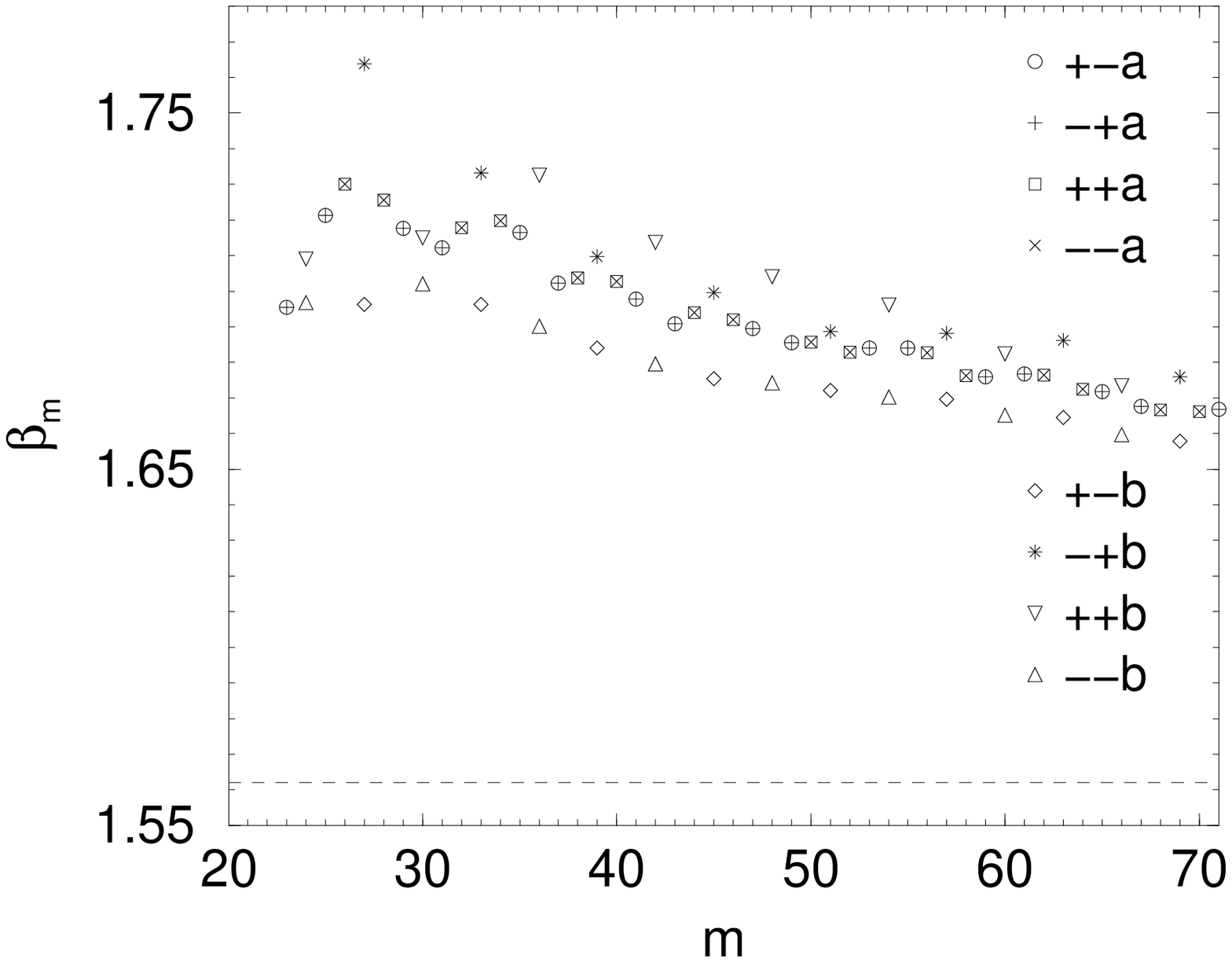}
\caption[]{\footnotesize $\beta_m$ vs. mode index $m$. Dashed line marks the
semiclassical prediction $\beta \approx 1.562$.} 
\label{fig:beta} 
\end{figure}

The quantization condition~(\ref{eq:qcondition}) explains the single-mode
spectrum in \figs~\ref{fig:sigma} and~\ref{fig:modes}(a). With $n=1.466$ the
mode spacing is $2\pi/3\sqrt{3}n 
\approx 0.8248$ in agreement with $\nu=0.8238$ obtained by fitting to the
numerical data. However, the shift $\beta \approx 1.562$ differs a bit from
the numerically obtained shift $m_0\approx 1.7516$. 
To understand this discrepancy we plot in \fig~\ref{fig:beta} the phase shifts
$\beta_m$ defined as $\beta$ computed from \equ~(\ref{eq:qcondition})
inserting the
numerically computed values of $\real{kR}$. Clearly, going towards the
semiclassical limit $m\to\infty$ leaves the crude fitting value $m_0\approx
1.7516$ in favour of a smaller value $\beta_\infty =
\lim_{m\to\infty} \beta_m$ closer to our semiclassical prediction
$\beta \approx 1.562$. 

\begin{figure}[ht]
\includegraphics[width=7.0cm,angle=0]{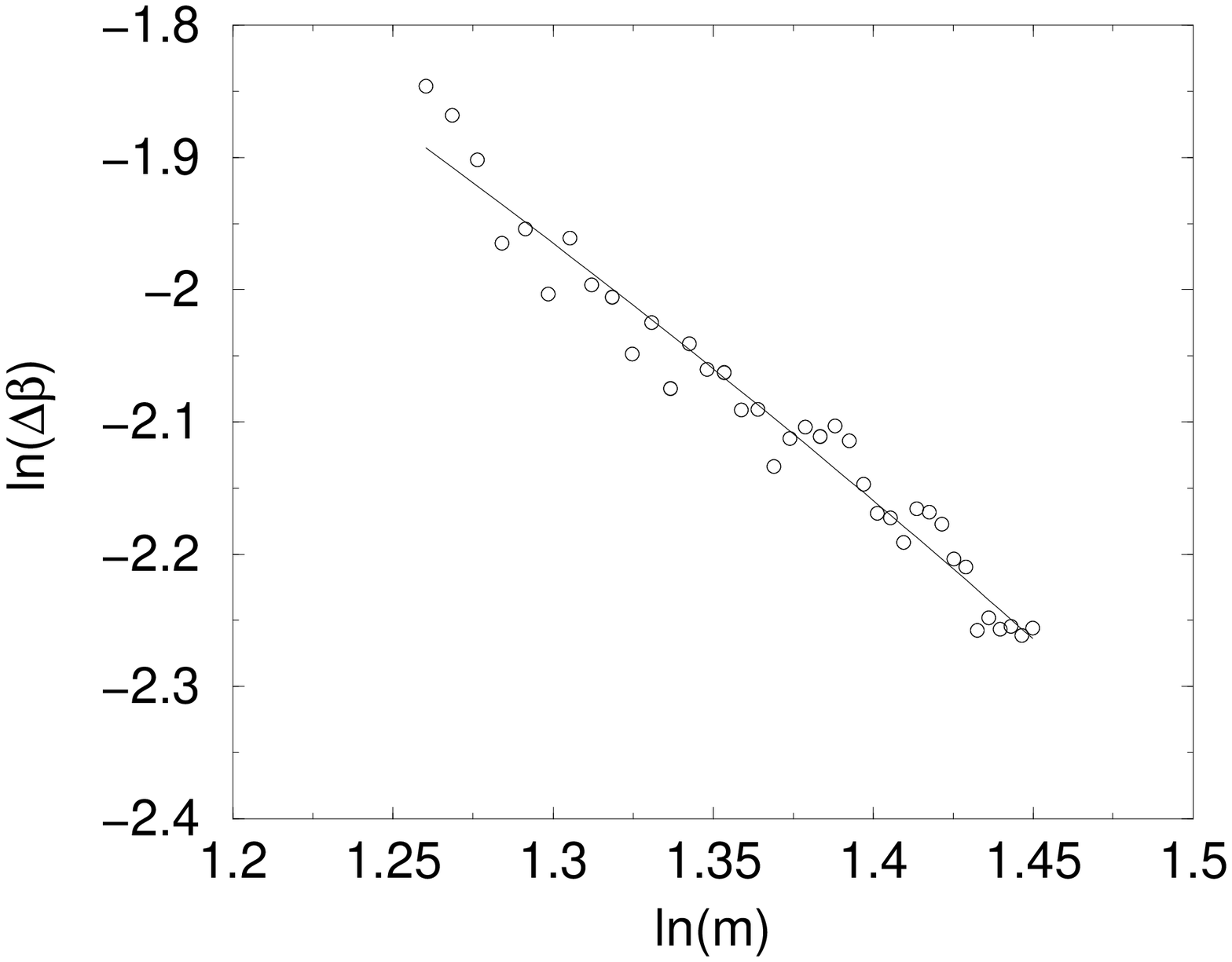}
\caption[]{\footnotesize $\ln{\Delta\beta}$ vs. $\ln{m}$ with $m\geq 34$. 
For quasi-degenerate modes the arithmetic mean of $\Delta\beta$ is taken.} 
\label{fig:betafit} 
\end{figure}

Let us assume that $\beta_m$ has the following form
\begin{equation}\label{eq:bm}
\beta_m = \gamma m^{-\delta}+\beta_\infty 
\end{equation}
and that $\beta_\infty$ equals our semiclassical solution
$\beta \approx 1.562$. 
Figure~\ref{fig:betafit} shows $\Delta\beta = \beta_m-\beta_\infty$ as
function of $m\geq 34$ in a log-log plot for fixed $n=1.466$. 
Linear regression gives $\gamma \approx 0.89$ and $\delta \approx
0.5038$. Based on this numerical finding we conjecture that the next order in
the semiclassical approximation~(\ref{eq:qcondition}) is of the form 
$\gamma/\sqrt{m}$. 

Observation of \figs~\ref{fig:lowresonance}(c) and (d) shows that 
the ``nodal lines'' close to the boundary of the cavity are not exactly
parallel to the boundary but slightly tilted (about $5^\circ$). This angular
shift decreases as one goes to higher-excited modes. For example, it is not
visible by eye in \fig~\ref{fig:highresonance}. 
Specular reflection of a plane wave at the boundary leads always to
nodal lines parallel to the boundary. Hence, it is reasonable to interpret 
the angular shift as a kind of {\it non-specular reflection}.
The angular shift can be traced back to the fact that the wave function
restricted to the boundary is a periodic function. Because of the periodicity,
the Fourier spectrum of the wave function along the boundary is
discrete. The discrete peaks can be related to a discrete set of allowed
angles of incidence. The angle $\theta_i = 60^\circ$ typically falls between
two neighboring discrete angles. In \sect~\ref{sec:angularshift} in the
Appendix we compute the angular shift analytically to first order as 
$\Delta\theta = \theta_i-60^\circ = \theta_i-\pi/3 \approx -\sqrt{3}\beta/m$
in agreement with the numerics (not shown). 
In the semiclassical limit $m\to\infty$ the angular shift vanishes. 
Our angular shift is different from those cases discussed in the context of
non-specular reflections (see e.g. Ref.~\cite{RBF73}), in that it happens above the
critical angle of total internal reflection. Moreover, it has nothing to do
with the angular Goos-H\"anchen effect (see e.g. Ref.~\cite{TDBFB95}), which is the
Goos-H\"anchen shift at a curved interface. 
\rem{
\begin{figure}[ht]
\includegraphics[width=6.5cm,angle=0]{dtheta.eps}
\caption[]{\footnotesize $\ln{\Delta\theta}$ vs. $\ln{m}$ for some $a$-modes.}
\label{fig:dtheta} 
\end{figure}
}

The number of wavelengths fitting on a path length determines the symmetry
class. Straightforward considerations give Tab.~\ref{tab:parities}. The 
semiclassical quantization procedure neither distinguishes between $+-$ and
$-+$ parties nor between $++$ and $--$ parties. Hence, within the semiclassical
approximation these modes are exactly degenerated, regardless whether they are
of type $a$ or $b$. This explains the numerical finding of quasi-degenerate
$b$-modes.
\begin{table}[!h]
\begin{center}
{
\begin{tabular}{|c|c|}\hline
$m$ & symmetry class \\ \hline
1 & $+-a$, $-+a$ \\
2 & $++a$, $--a$ \\
3 & $+-b$, $-+b$ \\
4 & $++a$, $--a$ \\
5 & $+-a$, $-+a$ \\
6 & $++b$, $--b$ \\ \hline
\end{tabular}
}
\caption[]{\label{tab:parities} \footnotesize Symmetry class and mode index $m$. The table is periodic in $m$ with period $6$.} 
\end{center}
\end{table}

The fact that our model predicts a single-mode spectrum, a one-parameter
$m=1,2,\ldots$ family of long-lived resonances, is related to the fact that
there is only one long-lived family of periodic rays (with short period). 
There is no ``transversal mode index'' which would be expected if the paraxial
approximation were applicable. Higher transverse modes correspond to rays with
angle of incidence considerably different from $60^\circ$. They are much 
shorter-lived which becomes clear in the following subsection when we discuss
the emission mechanisms. 

\subsection{Emission mechanisms}
Having derived a semiclassical quantization condition for the real part of
$kR$, we now compute its imaginary part.
The temporal behavior of the intensity of a resonant mode is $I\propto
\exp{[2\,\imag{\omega}t]}$. The outgoing relative intensity per unit time is 
\begin{equation}
\Delta I = -\frac{1}{I}\frac{dI}{dt} = -2\,\imag{\omega} = -2c\,\imag{k}\ .
\end{equation}
What are the mechanisms for this decay of intensity? Obviously, there is no
classical mechanism. In the framework of ray optics, the periodic rays in
\fig~\ref{fig:raymodel} remain forever in the cavity. Hence, we have to
include explicitly wave effects. 
We identify three candidates of such wave effects, all of which are related to
the corners. The first one is {\it diffraction at corners}. 
Corner diffraction may be a emission mechanism since a wave with finite
wavelength coming close to a corner is diffracted partly to the
exterior and partly back into the interior with another spectrum of directions
for which the condition of total internal reflection may not be fulfilled at
the next reflection. We will argue that corner diffraction is important for
the {\it emission directionality} but not for the {\it escape rate}. For the
escape rate we find two new relevant effects which are responsible for {\it
transport to corners}. We call them {\it boundary-wave leakage} and {\it
pseudointegrable leakage}.  

The boundary-wave leakage is illustrated in \fig~\ref{fig:raymodel}. 
An evanescent boundary wave travels along an infinitely extended dielectric
interface from $-\infty$ to $+\infty$. We assume that at a finite interface the
boundary wave fully separates from the interface at the corner. 
In \sect~\ref{sec:bwl} in the Appendix we determine the outgoing relative
intensity for any 
regular polygon (equilateral triangle, square, hexagon, ...) due to boundary
waves. In particular, for the hexagon we find 
\begin{equation}\label{eq:dibw}
\Delta I_{\text{bw}} =
\frac{3c}{4\,\real{k}R^2}\frac{n^3}{\sqrt{3n^2/4-1}(n^2-1)} \ .  
\end{equation}
\rem{
\begin{figure}[ht]
\includegraphics[width=7.0cm,angle=0]{finiteinterface.eps}
\caption{\footnotesize Escape of boundary waves at a corner.}
\label{fig:finiteinterface}
\end{figure}
}

The pseudointegrable leakage is due to the fact that wave optics does not
realize exactly $\theta_i = 60^\circ$ as already mentioned and shown in
\sect~\ref{sec:angularshift} in the Appendix.  
Putting this small angular deviation $\Delta\theta$ into the initial
conditions gives rise to 
rays with finite lifetime due to the pseudointegrable dynamics; see 
\fig~\ref{fig:raymodel}. In \sect~\ref{sec:pl} in the Appendix we estimate the 
outgoing relative intensity due to pseudointegrable leakage for
the hexagon as 
\begin{equation}\label{eq:dip}
\Delta I_{\text{p}} =
\frac{4\pi c}{3\real{k}R^2}\frac{\beta(n)}{n^2} \ .  
\end{equation}
The derivation can be easily extended to any regular polygon. Cavities with 
integrable internal dynamics, like the equilateral triangle and
the square, have $\Delta I_{\text{p}} = 0$. That means the neighborhood of
the long-lived family of periodic rays has roughly the same lifetime as
the periodic rays. In such a case, we expect a multimode spectrum in the
mesoscopic regime. Indeed, this has been found in scattering experiments
on the dielectric square in Ref.~\cite{PCC01}. 

\begin{figure}[ht]
\includegraphics[width=7.0cm,angle=0]{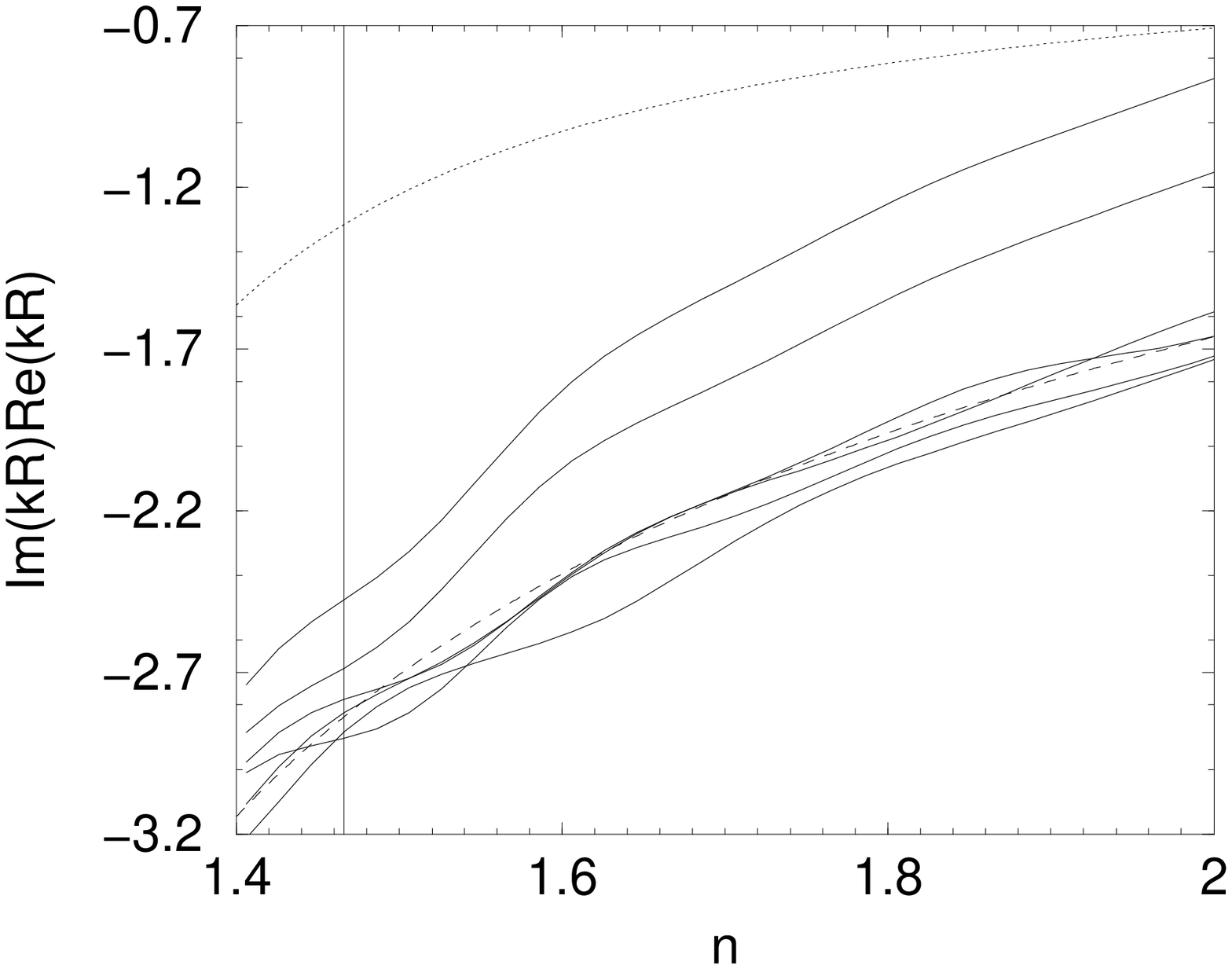}
\caption{\footnotesize $\imag{kR}\,\real{kR}$ vs. $n$ for several resonances
with $m\geq50$ (solid curves). The contribution of the boundary waves, 
\equ~(\ref{eq:fbw}), is shown as dotted curve. The entire contribution,
i.e. with pseudointegrable leakage, \equ~(\ref{eq:fp}), is shown as dashed
curve. The vertical line marks $n=1.466$.} 
\label{fig:dhigh} 
\end{figure}

Remarkably, both contributions in \eqs~(\ref{eq:dibw}) and (\ref{eq:dip}) have
the same dependence on $\real{k}$. In either case, the contribution vanishes
in the limit $\real{k}\to\infty$, reflecting their wave nature.
Adding both contributions, 
\begin{equation}
\Delta I = \Delta I_{\text{bw}} +\Delta I_{\text{p}} \ ,
\end{equation}
gives the central result
\begin{equation}\label{eq:kikr}
\imag{kR}\,\real{kR} = f(n) 
\end{equation}
with
\begin{equation}
f(n) = f_{\text{bw}}(n)+f_{\text{p}}(n) \ ,
\end{equation}
\begin{equation}\label{eq:fbw}
f_{\text{bw}}(n) = -\frac{3n^3}{8\sqrt{3n^2/4-1}(n^2-1)} \ ,
\end{equation}
and
\begin{equation}\label{eq:fp}
f_{\text{p}}(n) = -\frac{2\pi}{3}\frac{\beta(n)}{n^2} \ .
\end{equation}
For $n=1.466$ we find $\imag{kR}\,\real{kR} \approx -2.837$ in reasonable 
agreement with the numerical data in \figs~\ref{fig:complexplane} and
\ref{fig:modes}(b).  
Figure~\ref{fig:dhigh} compares the semiclassical result in
\eqs~(\ref{eq:kikr})-(\ref{eq:fp}) to the numerical data as function of
$n$. While some resonances are very well described by the semiclassical
approximation, some are only roughly described. The latter cases correspond to
the strong fluctuations around the semiclassical hyperbola~(\ref{eq:kikr}) 
already visible in \fig~\ref{fig:complexplane}.  

Our ray model also explains some aspects of the mode structure in
\figs~\ref{fig:highresonance} and \ref{fig:farfield}. The
whispering-gallery-like structure is related to the geometry of long-lived
rays; see \fig~\ref{fig:raymodel}. The envelopes of the emission direction are
described by the boundary-wave leakage and the pseudointegrable leakage;
cf. \figs~\ref{fig:highresonance}, \ref{fig:farfield}
and~\ref{fig:raymodel}. The distribution within the envelopes is not predicted
by our model. Here, corner diffraction is important because both the boundary
waves and the pseudointegrable ray dynamics lead to escape arbitrarily close
to the corner. 

Our ray model gives a natural explanation for the sensitivity to rounding of
corners. Let us assume that a sufficiently small rounding only slightly 
weakens the loss due to boundary waves, whereas it reduces corner diffraction 
strongly. Numerical simulations of ray dynamics in rounded hexagons show
that the periodic rays, see \fig~\ref{fig:raymodel}, are stabilized. This 
statement is independent on the particular chosen boundary parametrization.
Hence, there is no pseudointegrable leakage. This leads to considerably
larger lifetimes and emission parallel to the facets; compare 
\figs~\ref{fig:highresonance}-\ref{fig:roundedfar}. 

\section{Conclusion}
\label{sec:conclusion}
We have discussed the properties of long-lived resonances in hexagonally 
shaped dielectric microcavities. These microcavities play an important role for
experiments on a class of microcrystal lasers~\cite{VKLISLA98,BILNSSVWW00}.

The numerical analysis revealed the following facts: 
(i) The resonance positions in the complex plane are approximated by a 
hyperbola. This fact is
relevant for future experiments on microcrystal lasers because it implies that 
the laser threshold is lower for larger crystals. 
(ii) The near-field intensity pattern show a whispering-gallery-like 
structure.
(iii) The emission is at corners in agreement with the experiments.
(iv) The emission is directed. The high-intensity directions are not
parallel to the facets.   
(v) The lifetimes and the emission directionality are sensitive to
rounding of the corners. 

The numerical analysis does not cover the full experimental regime. To
overcome this limitation, we have introduced a semiclassical approximation 
which can be easily extended to any cavity of regular polygonal geometry. Our 
semiclassical ray model contains two new emission mechanisms: leakage due to
boundary waves and due to the pseudointegrable ray dynamics. Explicit 
expressions for the resonance positions can be given even though the system 
is not integrable. The semiclassical approximation explains all 
numerical findings (i)-(v) in an intuitive way, except the emission
directionality. To describe the directionality properly it is necessary to 
consider corner diffraction in the future.     

Our results are not only relevant for microcrystal lasers
but also for other kinds of polygonal-shaped microlasers. To name a few:
hexagonal lasers with attached optical 
waveguides based on semiconductor heterostructures~\cite{AKA95},
hexagonally shaped solid polymer dye microcavities~\cite{SLAFH00},
equilateral-triangular laser cavities fabricated
from semiconductor heterostructures~\cite{CKLHNXLPC00}, and square laser
micropillar cavities based on dye-doped polymers~\cite{CPC02}.

\begin{acknowledgments}
I would like to thank S. W. Kim, T. Harayama, H. Schomerus, M. Hentschel,
F. Laeri, and J.~U. N\"ockel for discussions. 
The work was supported by the Volkswagen foundation (project
``Molekularsieblaser-Konglomerate im Infraroten'').
\end{acknowledgments}

\begin{appendix}
\section{Angular shift}
\label{sec:angularshift}
In this section we discuss the semiclassical deviation of the angle of
incidence from $60^\circ$. Let us first investigate 
$--$ and $++$-modes. Along the boundary, these modes are periodic with
period $3R$, i.e. $\psi(s+3R) = \psi(s)$, where the arclength $s\in[0,6R)$
parametrizes the boundary. Hence, the wave function along the boundary can be  
expanded as the following Fourier series
\begin{equation}
\psi(s) = \sum_{M=-\infty}^{\infty}A_M\exp{\left(i\frac{2\pi}{3R}Ms\right)} \ .
\end{equation} 
From this we see that the conjugate momentum to $s$ or in other words the
momentum component parallel to the boundary is ``quantized'' according to
\begin{equation}
n\real{k}\sin\theta = \frac{2\pi}{3R}M
\end{equation}
with integer $M$. Inserting the quantization condition~(\ref{eq:qcondition})
leads to 
\begin{equation}
\sin\theta = \sqrt{3}\frac{M}{m+\beta} \ .
\end{equation}
Linearising this equation around $\theta\approx\theta_i = 60^\circ = \pi/3$
yield in the semiclassical regime $m\gg\beta$
\begin{equation}
\Delta\theta =
2\sqrt{3}\left(\frac{M}{m}-\frac{1}{2}\right)-2\sqrt{3}\frac{M}{m}\frac{\beta}{m}  \ .
\end{equation}
The smallest $\Delta\theta$ is realized for $M=m/2$ ($m$ is an even integer for
$--$ and $++$-modes, see \tab~\ref{tab:parities}), 
\begin{equation}\label{eq:Dt}
\Delta\theta = -\sqrt{3}\frac{\beta}{m} = -\frac{2\pi\beta}{3n\real{kR}} \ .
\end{equation} 
Analogous arguments concerning $-+$ and $+-$-modes give the same result as in
\equ~(\ref{eq:Dt}). 

\section{Boundary-wave leakage}
\label{sec:bwl}
In this section we compute the leakage due to boundary waves. To estimate the 
loss we consider the total 
internal reflection ($\theta_i > \theta_c$) of a plane wave at an infinitely
extended dielectric interface. This
consideration is justified if $\real{kR} \gg 1$ and $\theta_i$ not to close to
the critical angle for total internal reflection $\theta_c$. 
Since $\theta_e$ is a complex number in \equ~(\ref{eq:evanescent}), the wave
in the lower-index region is evanescent, i.e. it decays exponentially with
increasing distance from the boundary. Along the boundary the evanescent wave 
propagates with constant velocity. The corresponding total energy (or better
intensity) flux at a given point at the boundary can be computed by means of
\begin{equation}\label{eq:sigma}
\sigma = \int_0^\infty S_x dy \ ,
\end{equation}
where $S_x$ is the $x$-component of the Poynting vector~\cite{Jackson83} 
\begin{equation}\label{eq:poynting}
{\bf S} =  \frac{c}{8\pi} \real{{\bf E}\times{\bf H}^*} = -\frac{c}{8\pi k}\real{i\psi\nabla\psi^*} \ .
\end{equation}
Integration of \equ~(\ref{eq:sigma}) using \eqs~(\ref{eq:poynting}) and
~(\ref{eq:evanescent}) yields
\begin{equation}
\sigma = \frac{c}{4\pi}\frac{n\sin\theta_i}{k\sqrt{n^2\sin^2\theta_i-1}}
\frac{1}{1+\alpha^2} \ ,
\end{equation}
with $\alpha$ from \equ~(\ref{eq:alpha}).

\begin{figure}[ht]
\includegraphics[width=5.0cm,angle=0]{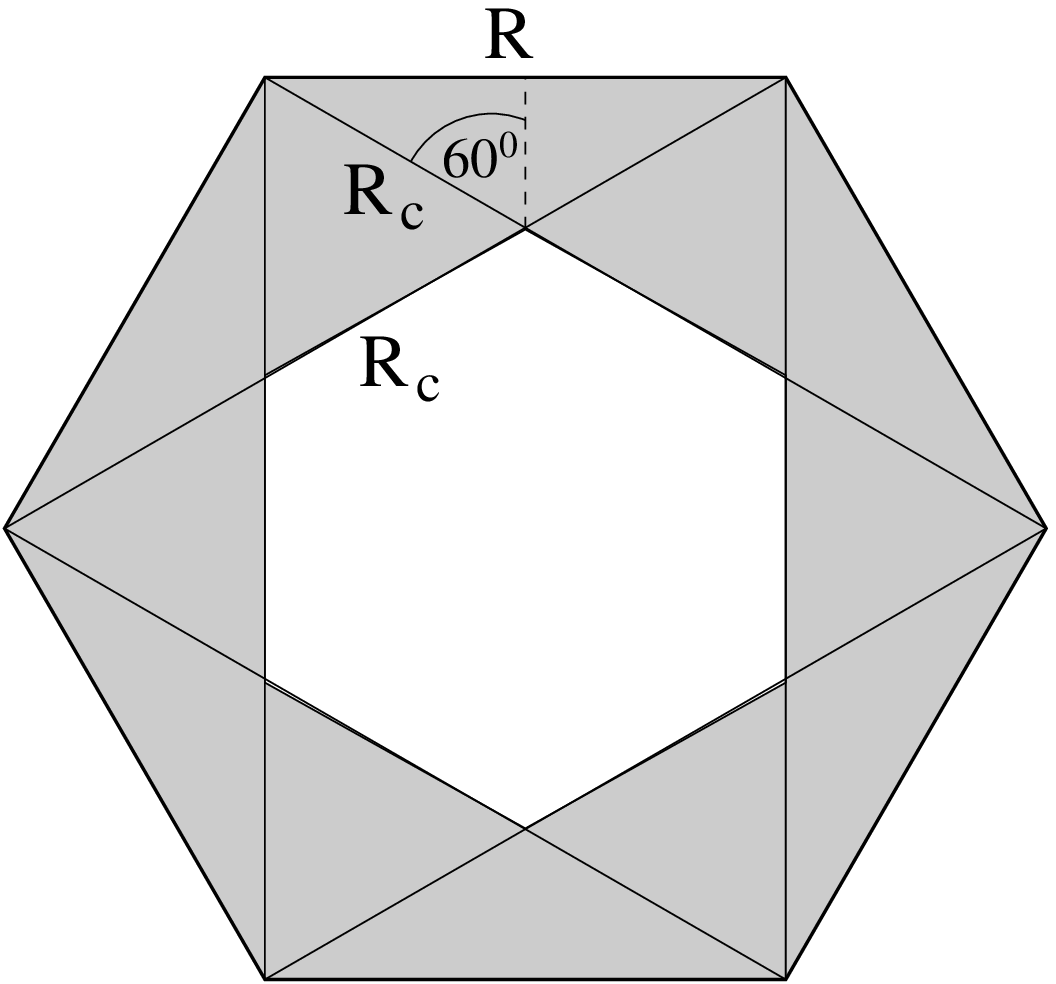}
\caption{\footnotesize Region (empty hexagon) not accessible by the family of
long-lived rays. The region is bounded by a hexagonal ``caustic'' with
side length $R_{\text{c}} = R/\sqrt{3}$.}
\label{fig:caustic} 
\end{figure}

Now we relate the total flux to the intensity inside the resonator. The
intensity of the wave function $\psi_i+\psi_r$ per area can be easily
calculated to be $1/4\pi$. Assuming that the boundary waves fully leave the
cavity at corners, the outgoing relative intensity per unit time is given by
\begin{equation}\label{eq:Ibw}
\Delta I_{\text{bw}} =
\frac{c}{A}\frac{n\sin\theta_i}{k\sqrt{n^2\sin^2\theta_i-1}} \frac{\chi}{1+\alpha^2} 
\ ,  
\end{equation}
where $\chi$ is the number of corners (where the light is emitted), $A$ is the
area of the resonator covered by the family of long-lived rays. 
Formula~(\ref{eq:Ibw}) is valid for any regular polygon. 
For the hexagon, $\chi = 6$ and $A$ is given by the area of the hexagon
minus the region not accessible by the family of long-lived rays 
as depicted in \fig~\ref{fig:caustic}. Elementary geometry yields 
\begin{equation}
A = \frac{3\sqrt{3}}2(R^2-R_{\text{c}}^2) = \sqrt{3}R^2 
\end{equation}
The final result for the hexagon is then given in \equ~(\ref{eq:dibw}).

\section{Pseudointegrable leakage}
\label{sec:pl}
Here we compute the pseudointegrable leakage. We put the angular shift 
$\Delta\theta$ from \equ~(\ref{eq:Dt}) into the initial conditions for
the ray dynamics. Elementary geometry, see \fig~\ref{fig:raymodel}, shows that
after one round trip, i.e. 6 bounces, the angular deviation gives rise to the
spatial deviation $\Delta s = |2l\Delta\theta|$. The time for each round trip
is $\Delta t = ln/c$. 

From \fig~\ref{fig:raymodel} it is clear that points on the boundary within
the distance $\Delta s$ from a corner leave the cavity after the next round
trip. Hence, the relative outgoing intensity can be computed as  
\begin{equation}
\Delta I_{\text{p}} = \frac{1}{R}\frac{\Delta s}{\Delta t} =
\frac{4\pi c}{3\real{k}R^2}\frac{\beta(n)}{n^2} \ .  
\end{equation}

\end{appendix}

\bibliographystyle{prsty}
\bibliography{../../bib/fg4,../../bib/extern}

\end{document}